\documentclass[prl,showpacs,showkeys,nofootinbib,twocolumn,floatfix]{revtex4}

\usepackage{ulem}
\usepackage{graphicx}  %
\usepackage{amsmath}
\usepackage{bm}  %

\newcommand{\bea}{\begin{eqnarray}}
\newcommand{\eea}{\end{eqnarray}}
\newcommand{\beq}{\begin{equation}}
\newcommand{\eeq}{\end{equation}}
\newcommand{\bqa}{\begin{eqnarray}}
\newcommand{\eqa}{\end{eqnarray}}

\def\mqo2{{\!\!\!}}

\begin{document}

\title{
The Avalanche Mechanism for Atom Loss \\
near an Atom-Dimer Efimov Resonance}

\author{Christian Langmack}
\affiliation{Department of Physics,
         The Ohio State University, Columbus, OH\ 43210, USA}

\author{D.~Hudson Smith}
\affiliation{Department of Physics,
         The Ohio State University, Columbus, OH\ 43210, USA}

\author{Eric Braaten}
\affiliation{Department of Physics,
         The Ohio State University, Columbus, OH\ 43210, USA}

\date{\today}

\begin{abstract}
An Efimov trimer near the atom-dimer threshold can increase 
the atom loss rate in ultracold trapped atoms through the
{\it avalanche mechanism} proposed by Zaccanti et al.
A 3-body recombination event creates an energetic atom and dimer, 
whose subsequent elastic collisions produce additional atoms 
with sufficient energy to escape from the trapping potential.
We use Monte Carlo methods to calculate the average number of atoms lost 
and the average heat generated by recombination events in both a 
Bose-Einstein condensate and a thermal gas.
We take into account the energy-dependence of the 
cross sections and the spatial structure of the atom cloud.  
We confirm that the number of atoms lost can be much larger than the 
naive value 3 if there is an Efimov trimer near the atom-dimer threshold.
This does not produce a narrow loss feature, but it can significantly 
affect the determination of Efimov parameters.
\end{abstract}

\smallskip
\pacs{34.50.-s, etc.}
\keywords{
Scattering of atoms and molecules, etc.}
\maketitle

Particles with short-range interactions and an S-wave scattering length $a$
that is large compared to the range have universal low-energy properties 
that depend on $a$ but not on other details of the interactions 
\cite{Braaten:2004rn}.
Universality provides connections between various fields of physics,
including atomic and molecular, condensed matter, nuclear, and particle physics.
This has stimulated theoretical progress in few-body physics.
Further stimulus has come from the use of ultracold trapped atoms 
to study reaction rates at extremely low energies.

Since particles with large scattering length are essentially indivisible 
at low energies, we will refer to them as {\it atoms}.
In the 2-atom sector, the universal properties for $a>0$  
include the existence of a {\it shallow dimer}
(weakly-bound diatomic molecule).
In many cases, including identical bosons, 
the universal properties in the 3-atom sector
include the existence of a sequence of universal triatomic molecules
called {\it Efimov trimers} \cite{Efimov70}.
In the zero-range limit, the spectrum of Efimov trimers 
is invariant under discrete scale transformations \cite{Efimov73}.
For identical bosons, the discrete scaling factor is approximately 22.7.
Reaction rates among three low-energy atoms also respect 
discrete scale invariance \cite{Efimov79}.
We refer to universal few-body phenomena with discrete scaling behavior 
as {\it Efimov physics}.

Ultracold trapped atoms provide an ideal laboratory for studying 
Efimov physics,  because $a$
can be controlled experimentally using Feshbach resonances.
The simplest probes of Efimov physics 
are loss features:  local maxima and minima in the atom loss rate 
as functions of $a$.
The most dramatic signature of an Efimov trimer is the resonant
enhancement of the 3-body recombination rate when
there is an Efimov trimer near the 3-atom threshold \cite{EGB-99}.
The first observation of such a loss feature in 
an ultracold gas of $^{133}$Cs atoms \cite{Grimm:06} 
revealed a line shape consistent with 
universal predictions \cite{Braaten:2003yc}.

In a mixture of atoms and shallow dimers, a narrow loss feature 
can also be caused by an Efimov trimer near the atom-dimer threshold.
We will refer to a scattering length $a_*$ for which an Efimov trimer
is exactly at the threshold as an {\it atom-dimer resonance}.
For $a$ near $a_*$, there is resonant enhancement of both the
elastic scattering of an atom and the shallow dimer  
and their inelastic scattering 
into an atom and a {\it deep dimer} (strongly-bound molecule).  
The large binding energy of the deep dimer gives
the outgoing atom and dimer large enough 
kinetic energies to escape from the trapping potential. 
The resulting peak in the atom loss rate near $a_*$ was first observed 
in a mixture of $^{133}$Cs atoms and dimers \cite{Grimm:0807}.

There have also been observations of enhanced loss rates 
near $a_*$ in systems consisting of atoms only.
Zaccanti et al.\ observed a narrow loss peak near the predicted position 
of an atom-dimer resonance in a Bose-Einstein condensate (BEC)
of $^{39}$K atoms \cite{Zaccanti:0904}.
They also observed a loss peak in a thermal gas 
near the next atom-dimer resonance, at a scattering length larger 
by a factor of about 22.7.
Pollack et al.\ observed a loss peak near the predicted position 
of an atom-dimer resonance in a BEC 
of $^{7}$Li atoms \cite{Hulet:0911}.
Machtey et al.\ observed such a loss peak 
in a thermal gas of $^{7}$Li atoms \cite{Khaykovich:1201}.
These loss features near the atom-dimer resonance are puzzling, 
because the equilibrium population of shallow dimers is expected to be 
negligible in these systems.

Zaccanti et al.\ proposed an {\it avalanche mechanism} 
for the enhancement of the atom loss rate near $a_*$ 
in systems consisting of atoms only \cite{Zaccanti:0904}.
The loss of atoms is initiated by a 3-body recombination event 
that produces an atom and a shallow dimer
with kinetic energies large enough to escape from the trap.
If they both escape, there would be 3 atoms lost. 
If the dimer instead scatters inelastically, 
the scattered atom is also lost.  
However the dimer can undergo multiple elastic collisions 
before ultimately escaping or suffering an inelastic collision,
and it may deliver enough energy to the scattered atoms 
to allow them to escape from the trap.
These atoms may also undergo multiple elastic collisions, 
resulting in an avalanche of additional lost atoms.
Near $a_*$, the resonant enhancement 
of the atom-dimer cross sections increases both the 
probability for the dimer to initiate an avalanche
and the probability for an inelastic collision.
The resulting increase in the number of atoms lost 
per recombination event could produce an observable loss feature.

In this paper, we present a quantitative analysis of the 
avalanche mechanism for atom loss.
We use Monte Carlo methods to generate avalanches of atoms initiated 
by recombination events with the appropriate probability distribution.
We calculate the average number of atoms lost 
and the average energy converted into heat from an avalanche.
We use the results to calculate the atom loss rate constant 
for both a Bose-Einstein condensate 
and a thermal gas of trapped atoms.

\begin{figure}[t]
\vspace*{-0.0cm}
\centerline{\includegraphics*[width=7cm,angle=0,clip=true]{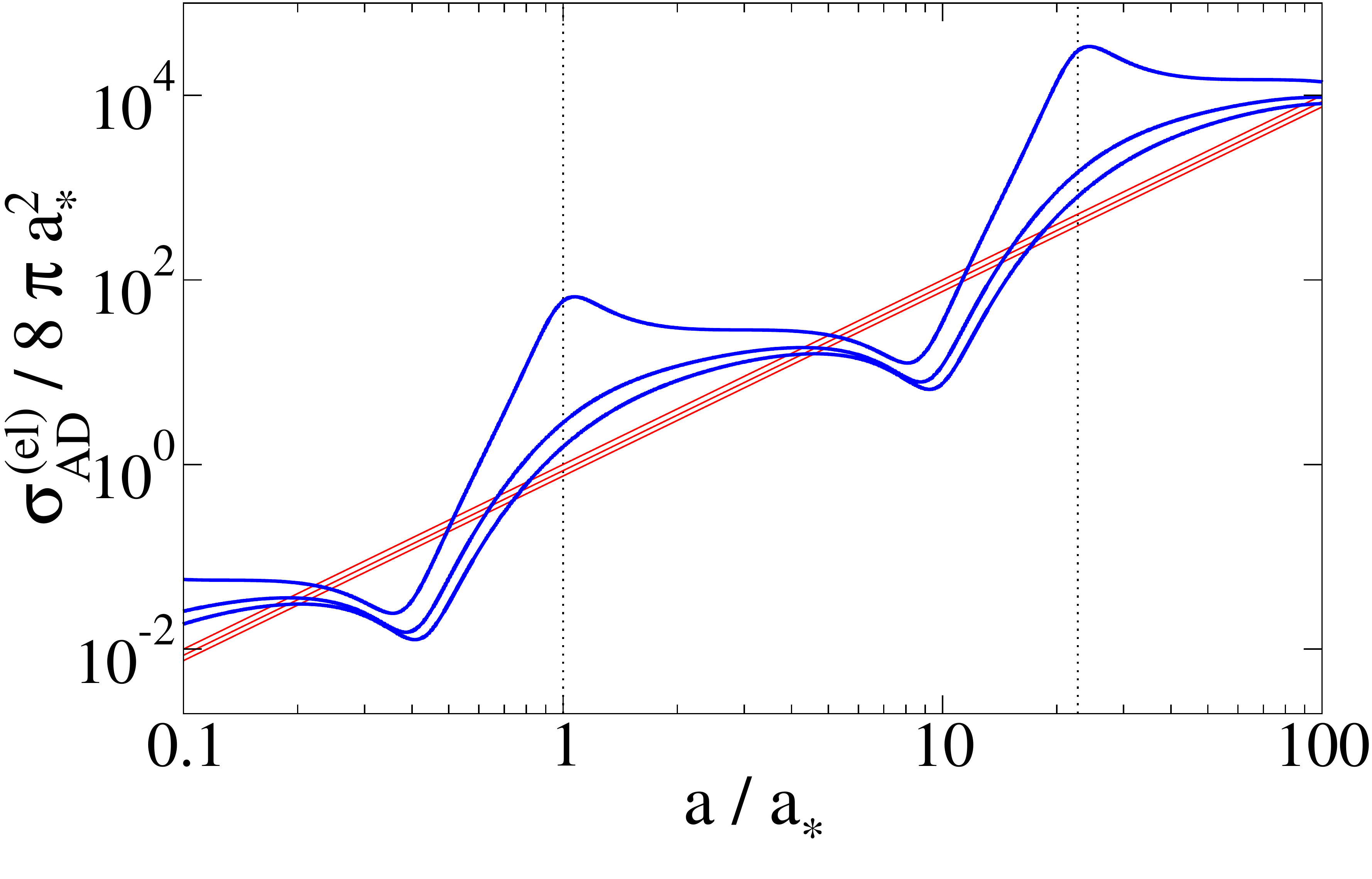}}
\vspace*{-0.0cm}
\centerline{\includegraphics*[width=7cm,angle=0,clip=true]{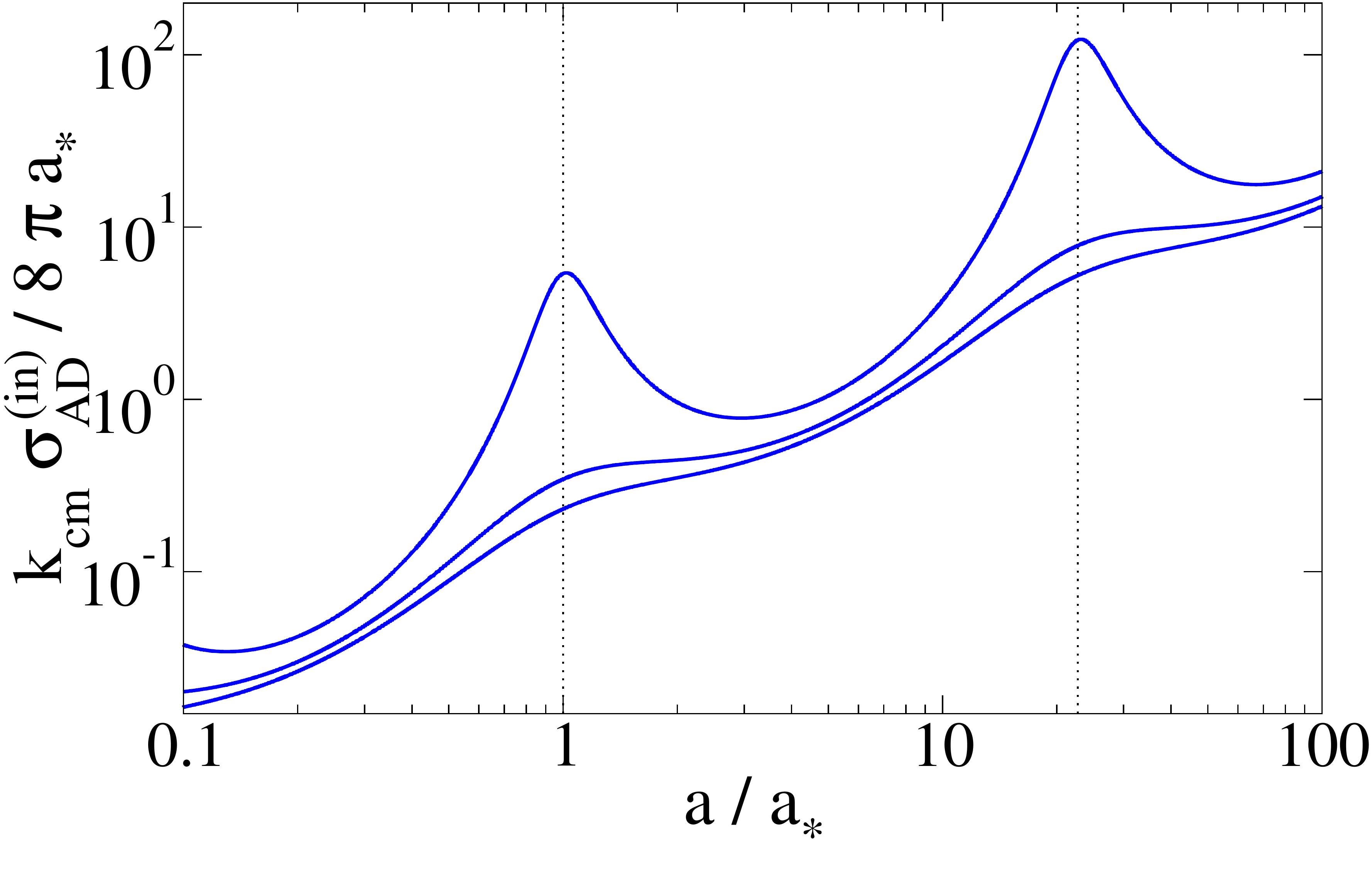}}
\vspace*{-0.0cm}
\caption{(Color online) 
Cross sections for elastic (upper panel) and inelastic (lower panel) 
atom-dimer collisions for $\eta_* = 0.2$.
The straight lines are elastic cross sections for atom-atom collisions.
The three curves for each case 
(in order of generally increasing cross sections)
are for the first scattering, a typical second scattering, 
and after many elastic scatterings.
The vertical dotted lines mark the positions of $a_*$ and $22.7~a_*$.
}
\label{fig:sigma}
\end{figure}

We consider identical bosons of mass $m$ 
with a large positive scattering length $a$.
The 2-body physics in our Monte Carlo model for the 
avalanche mechanism consists of the binding energy $E_d = \hbar^2/m a^2$
for the shallow dimer 
and the cross section $\sigma_{AA} = 8 \pi a^2/(1 + a^2 k_{\rm cm}^2)$ 
for elastic atom-atom scattering with center-of-mass 
wavenumber $k_{\rm cm}$.  The 3-body physics in our model 
consists of the rate constants $\alpha_{\rm shallow}$ and $\alpha_{\rm deep}$
for 3-body recombination at threshold into the shallow dimer 
and into deep dimers
and the cross sections $\sigma_{AD}^{\rm (el)}$ and $\sigma_{AD}^{\rm (in)}$
for elastic and inelastic atom-dimer scattering.
In the zero-range limit, these reaction rates are determined by
$a$ and two Efimov parameters:
the atom-dimer resonance $a_*$ and a dimensionless parameter $\eta_*$ 
that controls the decay width of an Efimov trimer \cite{Braaten:2003yc}. 
Analytic expressions for $\alpha_{\rm shallow}$ and $\alpha_{\rm deep}$
are given in Ref.~\cite{Braaten:2004rn}.
They are more conveniently expressed in terms of a parameter $a_{*0}$ 
that differs from $a_*$ by a universal ratio:
$a_{*0}/a_* = 4.4724$.
The 5 digits of accuracy in the ratio
are obtained by combining universal results
from Refs.~\cite{KHP:1001,Deltuva:1202}.
Parameterizations of $\sigma_{AD}^{\rm (el)}$ and 
$\sigma_{AD}^{\rm (in)}$ from the atom-dimer threshold
up to the dimer-breakup threshold
can be obtained from Ref.~\cite{Braaten:2004rn}.

We make several simplifying approximations in our model.
We ignore the effects of potential energies on the 
cross sections for the dimer and the atoms.
In the case of a BEC, we also ignore mean-field energies.
We also approximate the trajectories of the 
atoms and the dimer between collisions as straight lines.
In the first collision of the atom or dimer from the recombination event 
with a stationary atom, the center-of-mass wavenumber $k_{\rm cm}$ 
is $1/(\sqrt3 a)$ or $2/(3\sqrt3 a)$, respectively.
In the second collision with a stationary atom,
the typical $k_{\rm cm}$ is smaller by a factor of $1/\sqrt2$
for the atom and $\sqrt5/3$ for the dimer.
As the number of elastic collisions increases, $k_{\rm cm}$ decreases towards 0.
The universal cross sections for the first collision, 
a typical second collision, and after many elastic collisions 
($k_{\rm cm} \to 0$) are shown in Fig.~\ref{fig:sigma} for $\eta_* = 0.2$.
At $k_{\rm cm} = 0$, $\sigma_{AD}^{\rm (el)}$ 
and $k_{\rm cm} \sigma_{AD}^{\rm (in)}$
have dramatic peaks with maxima near $a_*$ and $22.7~a_*$. 
The elastic cross section also has deep minima near $0.38~a_*$ and $8.6~a_*$.  
For the first few collisions, $\sigma_{AD}^{\rm (el)}$ 
still has deep minima but there are no dramatic peaks.

The experimental inputs in our Monte Carlo model 
are the number $N_0$ of trapped atoms,
the frequencies $\nu_x$, $\nu_y$, and $\nu_z$
of the harmonic trapping potential,
the temperature $T$ of the atoms,
the scattering length $a$  (which can be controlled 
by varying the magnetic field near a Feshbach resonance), 
and the trap depth $E_{\rm trap}$, which
should be much larger than the energies of the trapped atoms.
Atoms and dimers that reach the edge of the atom cloud 
are assumed to be lost if their energies exceed 
$E_{\rm trap}$ and $2 E_{\rm trap}$, respectively.
The role of the remaining experimental inputs 
is to determine the number density $n(x,y,z)$ of the trapped atoms.
We consider two simple cases:
a BEC of atoms at zero temperature 
in the Thomas-Fermi limit and
a thermal gas of atoms above the critical temperature for BEC.
The rate at which the number $N$ of atoms in a thermal gas
decreases due to 3-body recombination is
\begin{equation}
dN/dt = 
- \left( \langle N_{\rm lost} \rangle \alpha_{\rm shallow} 
        + 3 \alpha_{\rm deep} \right) \langle n^2 \rangle N,
\label{dNdt}
\end{equation}
where $\langle n^2 \rangle$ and $\langle N_{\rm lost} \rangle$
are spatial averages weighted by $n(\bm{r})$ and $n^3(\bm{r})$,
respectively. The right side must be multiplied by $1/3!$ 
if the system is a BEC.
Atoms produced by the avalanche 
that have energy less than $E_{\rm trap}$
can never escape from the trapping potential and their kinetic energy 
will ultimately be transformed into heat.
The recombination heating rate in a thermal gas is 
$\langle E_{\rm heat} \rangle \alpha_{\rm shallow} \langle n^2 \rangle N$,
where $\langle E_{\rm heat} \rangle$ 
is the average heat from a single avalanche.

The development of an avalanche can be decomposed into discrete steps
corresponding to the recombination event and the subsequent scattering events.
For each event, 
the subsequent state of the avalanche has a simple probability distribution.  
(a)~The position $(x,y,z)$ of the recombination point
has a distribution proportional to $n^3(x,y,z)$.
(b)~The momenta of the outgoing particles from an event
have a distribution that is isotropic in the center-of-mass frame. 
(c)~An atom or dimer flies beyond 
the edge of the atom cloud with probability 
$\exp(- \sigma \int \!n\, \hbox{d}\ell)$,
where $\sigma$ is $\sigma_{AA}$ for an atom
and $\sigma_{AD}^{\rm (el)} + \sigma_{AD}^{\rm (in)}$  for a dimer
and where $\int \!n\, \hbox{d}\ell$ is the column density
integrated from the position of the previous collision
out to infinity along a straight path in the direction of the momentum.
If the random number determines that an atom or dimer 
fails to reach the edge of the atom cloud, it is also used to  
determine the position where it scatters.
(d)~Given that a dimer scatters, it scatters inelastically with
probability
$\sigma_{AD}^{\rm (in)}/(\sigma_{AD}^{\rm (el)} + \sigma_{AD}^{\rm (in)})$.
All these simple probability distributions together determine 
the probability distribution of avalanches.

We generate avalanches with the appropriate probability distribution 
using a Monte Carlo method that produces a binary tree
whose nodes represent events.
The branches represent the two outgoing particles from each event.
There are also terminal nodes that correspond to atoms and dimers
whose ultimate fate is determined.
The conditions for a terminal node 
depend on the kinetic energy $E$ of the particle.
(a)~If an atom has $E < E_{\rm trap}$,
it remains trapped.
(b)~If an atom that reaches the edge of the atom cloud
has $E > E_{\rm trap}$, it is lost.
(c)~If a dimer has an inelastic collision, 
both it and the scattered atom are lost.
(d)~If a dimer that reaches the edge of the cloud
has $E > 2 E_{\rm trap}$, it is lost.
(e)~If a dimer that reaches the edge of the cloud
has $E < 2 E_{\rm trap}$,
it will return to the cloud 
and eventually suffer an inelastic collision.
The terminal nodes give contributions to the number of atoms lost
and to the heat of the remaining atoms.
Adding these contributions from all the terminal nodes, we get 
$N_{\rm lost}$ and $E_{\rm heat}$ for the avalanche.
We calculate $\langle N_{\rm lost} \rangle$ and 
$\langle E_{\rm heat} \rangle$ by averaging over many 
avalanches.  More than 100,000 avalanches
are sometimes required to get smooth results for
$\langle N_{\rm lost} \rangle$ and $\langle E_{\rm heat} \rangle$ 
as functions of $a$.

Zaccanti et al.\ developed a simple probabilistic model 
for the avalanche process \cite{Zaccanti:0904}.
In the {\it Zaccanti model}, the avalanche is reduced to a
discrete sequence of dimer scattering events.
A variable number of elastic collisions is followed either by the escape 
of the dimer from the trap or by a final inelastic collision.
There is one lost atom for each elastic collision up to a maximum 
number that is determined by the trap depth $E_{\rm trap}$.
The relative probability for each sequence of scattering events 
is determined by the mean column density
and by $\sigma_{AD}^{\rm (el)}$ and $k_{\rm cm} \sigma_{AD}^{\rm (in)}$.
The Zaccanti model is greatly simplified in several ways 
compared to our Monte Carlo model:
(a)~The spatial structure of the avalanche is ignored.
(b)~Elastic scattering of the atoms is not considered.
(c)~The energy dependence of 
$\sigma_{AD}^{\rm (el)}$ and $k_{\rm cm} \sigma_{AD}^{(\rm in)}$ 
is not taken into account.
Zaccanti et al.\ used their model to calculate 
$\langle N_{\rm lost} \rangle$ for their experiment 
with $^{39}$K atoms \cite{Zaccanti:0904}.
It predicts that $\langle N_{\rm lost} \rangle$
increases from its background value of 3 to about 13 
near the atom-dimer resonance.  
The resulting prediction for the atom loss rate agrees 
qualitatively with the loss feature they observed near $30.4~a_0$.

Machtey et al.\ developed an alternative probabilistic model 
for the avalanche process \cite{Khaykovich:1111}.
They made the same simplifications as in
the Zaccanti model, but they used different probabilities 
for the sequences of scattering events. 
Machtey et al.\  did not introduce the trap depth $E_{\rm trap}$,  
so they could not calculate $\langle N_{\rm lost} \rangle$.
Instead they calculated 
the average number $\bar N$ of dimer collisions.
They suggested that the loss feature near $a_*$ might be 
associated with the maximum of $\bar N$.

\begin{figure}[t]
\vspace*{-0.0cm}
\centerline{\includegraphics*[width=5cm,angle=270,clip=true]{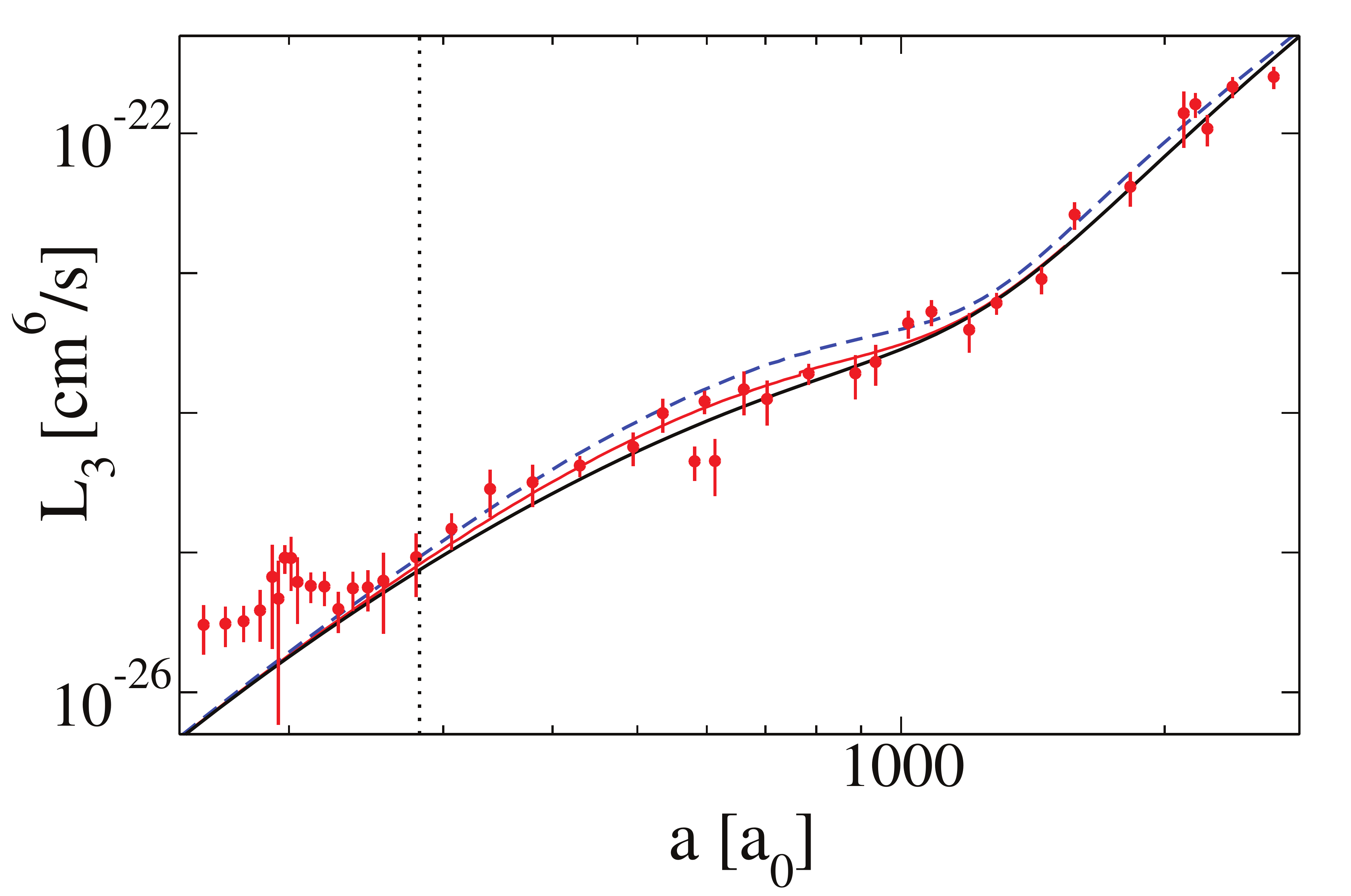}}
\vspace*{-0.0cm}
\caption{(Color online) 
Rate constant $L_3$ for 3-body recombination in $^7$Li atoms. 
The data are for the thermal gas from Ref.~\cite{Khaykovich:1201}.
The lower solid curve is the naive result for $\langle N_{\rm lost} \rangle = 3$
with $a_* = 282~a_0$ and $\eta_* = 0.2$.
The upper solid (red) and dashed (blue) curves are predictions 
of our Monte Carlo model 
for the thermal gas in Ref.~\cite{Khaykovich:1201}
and for the BEC in Ref.~\cite{Hulet:0911}.
The vertical dotted line marks the position of $a_*$.
}
\label{fig:L3}
\end{figure}

In the experiment with $^{39}$K atoms in Ref.~\cite{Zaccanti:0904},
the loss feature near $30.4~a_0$ is at a scattering length that 
may be too small for universal predictions to be reliable.
We therefore focus on the experiments with $^7$Li atoms.
In Ref.~\cite{Hulet:0911},
the data near the atom-dimer resonance  were obtained using a BEC 
of $^7$Li atoms in the $|1,+1\rangle$ hyperfine state
with $E_{\rm trap} \approx 0.5~\mu$K.
They measured the rate constant 
$L_3 = 3 (\alpha_{\rm shallow} + \alpha_{\rm deep})$ as a function of $a$.
The best fit to $L_3$ near an interference minimum
gave the Efimov parameters $a_{*0} = 2672~a_0$ and $\eta_* = 0.039$.
The accuracy of the determination of $a$ 
as a function of the magnetic field has been improved
and a reanalysis of that data 
is underway \cite{Hulet:1205}.
The parameter $a_{*0}$ will change significantly, but $\eta_*$ will not.
In Ref.~\cite{Khaykovich:1003}, $L_3$ was measured as a function of $a$ using 
a thermal gas of $^7$Li atoms in the $|1,+1\rangle$ hyperfine state 
with $T \approx 1.4~\mu$K and $E_{\rm trap} \approx 7~\mu$K. 
The data are shown in Fig.~\ref{fig:L3}.
The best fit to $L_3$ gives $a_{*0} = 1260~a_0$ and $\eta_* = 0.188$.
Using the universal ratio for $a_{*0}/a_*$, we obtain the prediction 
$a_* = 282~a_0$.  The curve for $a_* = 282~a_0$ and $\eta_* = 0.2$
shown in Fig.~\ref{fig:L3} provides a good fit over most of the range.
The data in Fig.~\ref{fig:L3} for $a$ below $220~a_0$ 
were presented in Ref.~\cite{Khaykovich:1201},
revealing the narrow loss feature near $200~a_0$.

\begin{figure}[t]
\vspace*{-0.0cm}
\centerline{\includegraphics*[width=7cm,angle=0,clip=true]{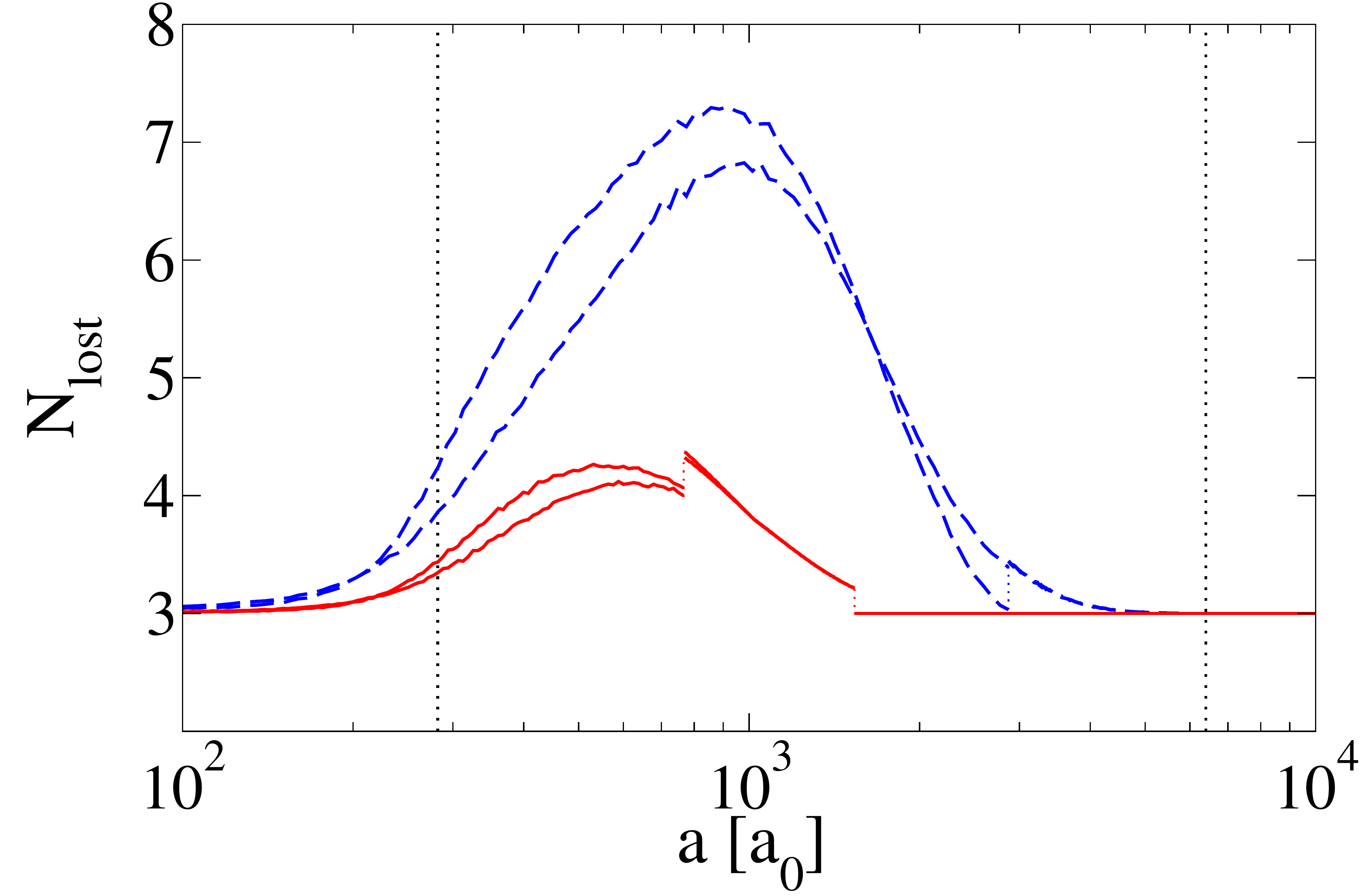}}
\vspace*{-0.0cm}
\centerline{\includegraphics*[width=7cm,angle=0,clip=true]{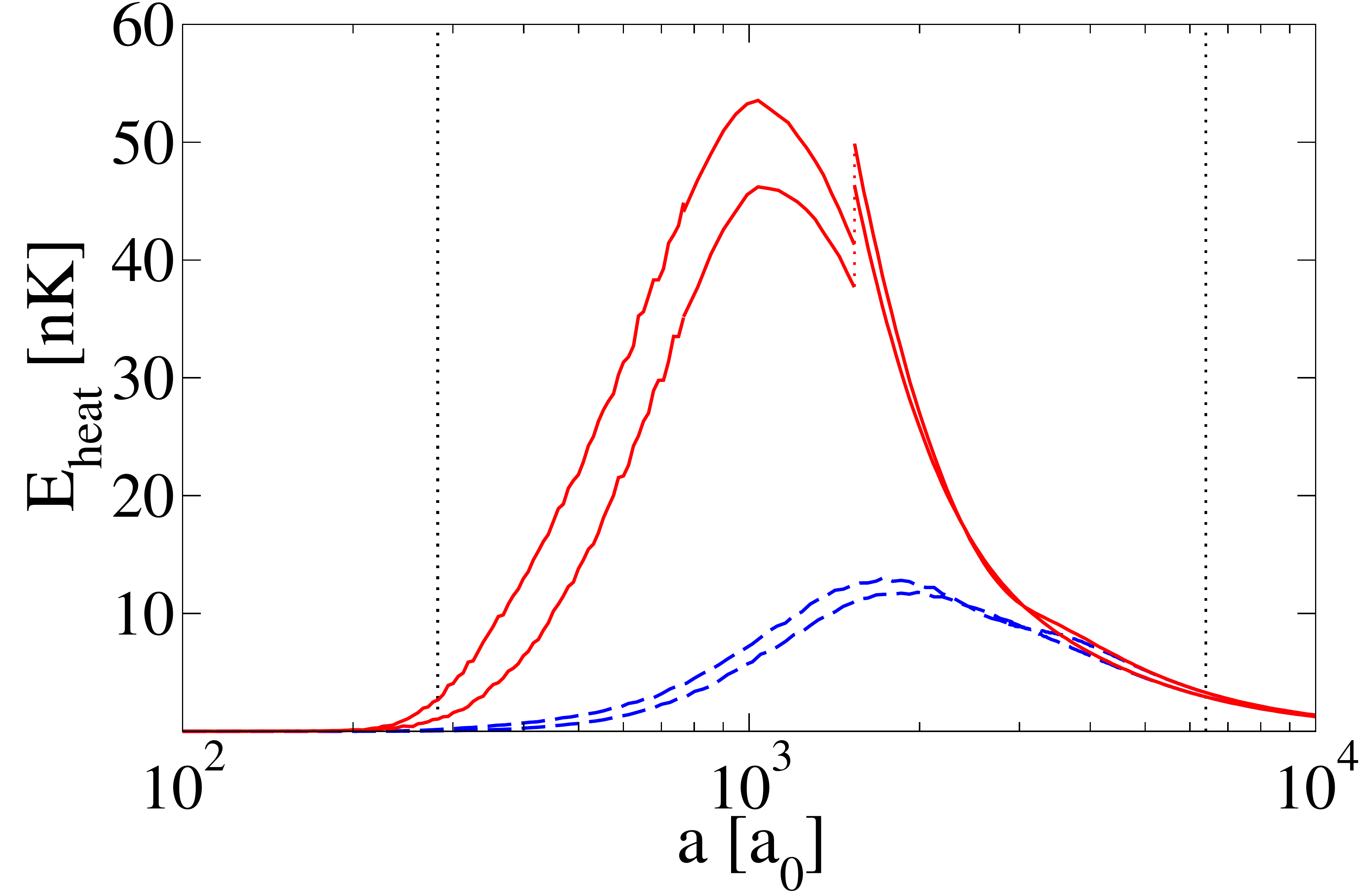}}
\vspace*{-0.0cm}
\caption{(Color online) 
Average number $\langle N_{\rm lost} \rangle$ of atoms lost (upper panel)
and average heat $\langle E_{\rm heat} \rangle$ generated (lower panel)
by an avalanche from 3-body recombination in $^7$Li atoms. 
The dashed (blue) and solid (red) curves are the predictions 
of our Monte Carlo model with $a_* = 225~a_0$ and 
$\eta_* = 0.2$ (lower curve) or $\eta_* = 0.04$ (higher curve)
for the BEC in Ref.~\cite{Hulet:0911} and
for the thermal gas in Ref.~\cite{Khaykovich:1201}.
The vertical dotted lines mark the positions of $a_*$ and $22.7~a_*$.
}
\label{fig:Nlost}
\end{figure}

In Fig.~\ref{fig:Nlost}, we show the predictions of our Monte Carlo model 
for $\langle N_{\rm lost} \rangle$ and $\langle E_{\rm heat} \rangle$ 
as functions of $a$  for both $^7$Li experiments.
We set $a_* = 282~a_0$ 
and we consider both $\eta_* = 0.2$ and $\eta_* = 0.04$.
There are discontinuities in $\langle N_{\rm lost} \rangle$ 
and $\langle E_{\rm heat} \rangle$ when the energies of the 
atom and the dimer from the recombination event are equal to 
$E_{\rm trap}$ and $2 E_{\rm trap}$, respectively.
Both $\langle N_{\rm lost} \rangle$ and $\langle E_{\rm heat} \rangle$ 
have broad peaks whose maxima occur well above $a_*$.
This is because the first few
elastic collisions of the dimer from the recombination event 
are the most important for generating an avalanche.
As shown in Fig.~\ref{fig:sigma}, the maxima in the elastic cross sections 
for the first few collisions occur well above $a_*$.
Decreasing $\eta_*$ by a factor of 5 does not give dramatic changes 
in $\langle N_{\rm lost} \rangle$ and $\langle E_{\rm heat} \rangle$.

If the avalanche mechanism is taken into account, as in Eq.~(\ref{dNdt}),
the rate constant is 
$L_3 = \langle N_{\rm lost} \rangle \alpha_{\rm shallow} 
+ 3 \alpha_{\rm deep}$.
In Fig.~\ref{fig:L3}, we show the predictions for both experiments
for $L_3$ as a function of $a$ 
using $a_* = 282~a_0$ and $\eta_* = 0.2$.
If $\eta_*$ were decreased to 0.04, 
the changes in the curves would be obvious only near 
$a_{*0} \approx 1260~a_0$, where the minimum would be 
about a factor of 5 below the data.
There is no narrow loss feature near $a_*$, but
instead there is a broad enhancement in $L_3$ 
in the region between $a_*$ and $a_{*0}$.
The enhancement is large enough that it could affect the fitted values 
of the Efimov parameters.
We are unable to get a narrow loss feature like that in the 
data in Fig.~\ref{fig:L3} for any values of the parameters.

Our results for the avalanche mechanism also suggest that 
the effects of heating should be taken into account differently
in the analysis of atom loss data at positive scattering lengths.
In addition to the recombination heating associated with $E_{\rm heat}$,
one must take into account the disappearance of the three atoms 
that undergo recombination \cite{Grimm:03}.
In a thermal gas, the resulting rate of change in the temperature is
\begin{eqnarray}
dT/dt = 
\big( 
\left[ \langle E_{\rm heat} \rangle/(3 kT)
      + \langle N_{\rm lost} \rangle - 2 \right] \alpha_{\rm shallow} 
\nonumber
\\
+ \alpha_{\rm deep} 
\big) \langle n^2 \rangle T.
\label{dTdt}
\end{eqnarray}
This has the same dependence on $T$ as that assumed in Ref.~\cite{Grimm:03}.
In the coefficient of $\alpha_{\rm shallow}$,
the $1/kT$ term is determined by $\langle E_{\rm lost} \rangle$ 
and the additive constant $\langle N_{\rm lost} \rangle -2$
can differ from the naive value 1.
The effects of heating are usually taken into account by 
using the coupled equations for $dN/dt$ and $dT/dt$ to 
extrapolate to the initial value of $dN/dt$~\cite{Grimm:03}.
The $\langle N_{\rm lost} \rangle$ and 
$\langle E_{\rm heat} \rangle$ terms in Eqs.~(\ref{dNdt}) and (\ref{dTdt})
may have a significant effect on this extrapolation.
 
We have found that the avalanche mechanism does not produce 
a narrow loss feature near the atom-dimer resonance, 
but instead a broad enhancement of $L_3$ between $a_*$ and $a_{*0}$.
This can have a significant effect on the determination 
of Efimov parameters from data at positive scattering lengths.
The heating from the avalanche mechanism can also be 
important in the extrapolation to the initial atom loss rate.
All experiments on Efimov loss features at positive $a$ should 
probably be reanalyzed to take into account the effects of the 
avalanche mechanism.  If the observed loss features near $a_*$  
in $^{39}K$ and $^7$Li atoms survive such a reanalysis, 
the mechanism for these loss features will remain a puzzle.

\begin{acknowledgments}
We thank R.~Hulet, L.~Khaykovich, and M.~Zaccanti 
for valuable comments.
This research was supported in part by a joint grant from 
the Army Research Office 
and the Air Force Office of Scientific Research.
\end{acknowledgments}

\newpage


\begin{thebibliography}{99}

\bibitem{Braaten:2004rn}
  E.~Braaten and H.-W.~Hammer,
  Phys.\ Rept.\  {\bf 428}, 259 (2006).

\bibitem{Efimov70}
V.~Efimov,
Phys.\ Lett.\ {\bf 33B}, 563 (1970).

\bibitem{Efimov73}
V.~Efimov,
Nucl.\ Phys.\ A {\bf 210}, 157 (1973).

\bibitem{Efimov79}
V.~Efimov,
Sov.\ J.\ Nucl.\ Phys.\ {\bf 29}, 546 (1979).

\bibitem{EGB-99}
B.D.~Esry, C.H.~Greene, and J.P.~Burke,
Phys.\ Rev.\ Lett.\ {\bf 83}, 1751 (1999).

\bibitem{Grimm:06}
T.~Kraemer et al.,
Nature {\bf 440}, 315 (2006).

\bibitem{Braaten:2003yc}
  E.~Braaten and H.-W.~Hammer,
   Phys.\ Rev.\  A {\bf 70}, 042706 (2004).

\bibitem{Grimm:0807}
S.~Knoop et al.,
Nature Physics {\bf 5}, 227 (2009).

\bibitem{Zaccanti:0904}
M.~Zaccanti et al., 
Nature Physics {\bf 5}, 586 (2009). 
    
\bibitem{Hulet:0911}
S.E.~Pollack, D.~Dries, and R.G.~Hulet,
Science {\bf 326}, 1683 (2009).

\bibitem{Khaykovich:1201}
O.~Machtey et al.,
arXiv:1201.2396.
  
\bibitem{KHP:1001}
K.~Helfrich, H.-W.~Hammer, and D.S.~Petrov,
Phys.\ Rev.\ {\bf A81}, 042715 (2010). 
   
\bibitem{Deltuva:1202}
A.~Deltuva, 
arXiv:1202.0167. 

\bibitem{Khaykovich:1111}
O.~Machtey, D.A.~Kessler, and L.~Khaykovich,
Phys.\ Rev.\ Lett.\ {\bf 108}, 130403 (2012). 

\bibitem{Hulet:1205}
R.G.~Hulet, private communication.
  
\bibitem{Khaykovich:1003}
N.~Gross, Z.~Shotan, S.~Kokkelmans, and L.~Khaykovich,
Phys.\ Rev.\ Lett.\ {\bf 105}, 103203 (2010).

\bibitem{Grimm:03}
T.~Weber et al., 
Phys.\ Rev.\ Lett.\ {\bf 91}, 123201 (2003).

\end{thebibliography}
\end{document}